\documentclass{article}

\usepackage{graphicx}
\usepackage[square,numbers]{natbib}

\usepackage[preprint]{neurips_2023}

\usepackage[utf8]{inputenc} 
\usepackage[T1]{fontenc}    
\usepackage{hyperref}       
\usepackage{url}            
\usepackage{booktabs}       
\usepackage{amsfonts}       
\usepackage{nicefrac}       
\usepackage{microtype}      
\usepackage{xcolor}         

\title{OpenVoice: Versatile Instant Voice Cloning}

\author{%
  Zengyi Qin \thanks{Corresponding email: qinzy@mit.edu}\\
  MIT \& MyShell.ai \\
  \And 
  Wenliang Zhao \\
  Tsinghua University \\
  \And 
  Xumin Yu \\
  Tsinghua University \\
  \And 
  Xin Sun \\
  MyShell.ai
}

\begin{document}

\maketitle

\begin{abstract}
  We introduce OpenVoice, a versatile instant voice cloning approach that requires only a short audio clip from the reference speaker to replicate their voice and generate speech in multiple languages. OpenVoice represents a significant advancement in addressing the following open challenges in the field: \textbf{1) Flexible Voice Style Control.} OpenVoice enables granular control over voice styles, including emotion, accent, rhythm, pauses, and intonation, in addition to replicating the tone color of the reference speaker. The voice styles are not directly copied from and constrained by the style of the reference speaker. Previous approaches lacked the ability to flexibly manipulate voice styles after cloning. \textbf{2) Zero-Shot Cross-Lingual Voice Cloning.} OpenVoice achieves zero-shot cross-lingual voice cloning for languages not included in the massive-speaker training set. Unlike previous approaches, which typically require extensive massive-speaker multi-lingual (MSML) dataset\footnote{We use the term \textbf{MSML dataset} to denote a speech dataset with multiple languages, and each language has a vast amount of speakers. It is a typical requirement for training cross-lingual instant voice cloning models.} for all languages, OpenVoice can clone voices into a new language without any massive-speaker training data for that language. OpenVoice is also \textbf{computationally efficient}, costing tens of times less than commercially available APIs that offer even inferior performance. To foster further research in the field, we have made the source code\footnote{https://github.com/myshell-ai/OpenVoice} and trained model publicly accessible. We also provide qualitative results in our demo website\footnote{https://research.myshell.ai/open-voice}. OpenVoice has served more than 2M users worldwide as the voice engine of MyShell.ai.
\end{abstract}

\section{Introduction}

Instant voice cloning (IVC) in text-to-speech (TTS) synthesis means the TTS model can clone the voice of any reference speaker given a short audio sample without additional training on the reference speaker. It is also referred to as Zero-shot TTS. IVC enables the users to flexibly customize the generated voice and exhibits tremendous value in a wide variety of real-world applications, such as media content creation, customized chatbots, and multi-modal interaction between humans and computers or large language models. 

An abundant of previous work has been done in IVC. Examples of auto-regressive approaches include VALLE~\cite{wang2023neural} and XTTS~\cite{coquixtts}, which extract the acoustic tokens or speaker embedding from the reference audio as a condition for the auto-regressive model. Then the auto-regressive model sequentially generate acoustic tokens, which are then decoded to raw audio waveform. While these methods can clone the tone color, they do not allow users to flexibly manipulate other important style parameters such as emotion, accent, rhythm, pauses and intonation. Also, auto-regressive models are relatively computationally expensive and has relatively slow inference speed. Examples of non-autoregressive approach include YourTTS~\cite{casanova2022yourtts} and the recently developed Voicebox~\cite{le2023voicebox}, which demonstrate significantly faster inference speed but are still unable to provide flexible control over style parameters besides tone color. Another common disadvantage of the existing methods is that they typically require a huge MSML dataset in order to achieve cross-lingual voice clone. Such combinatorial data requirement can limit their flexibility to include new languages. In addition, since the voice cloning research~\cite{le2023voicebox, wang2023neural} by tech giants are mostly closed-source, there is not a convenient way for the research community to step on their shoulders and push the field forward.

We present OpenVoice, a flexible instant voice cloning approach targeted at the following key problems in the field:
\begin{itemize}
    \item In addition to cloning the tone color, how to have flexible control of other important style parameters such as emotion, accent, rhythm, pauses and intonation? These features are crucial for generating in-context natural speech and conversations, rather than monotonously narrating the input text. Previous approaches~\cite{casanova2022yourtts, coquixtts, wang2023neural} can only clone the monotonous tone color and style from the reference speaker but do not allow flexible manipulation of styles.
    \item How to enable zero-shot cross-lingual voice cloning in a simple way. We put forward two aspects of zero-shot capabilities that are important but not solved by previous studies:
    \begin{itemize}
        \item If the language of the reference speaker is not presented in the MSML dataset, can the model clone their voice?
        \item If the language of the generated speech is not presented in the MSML dataset, can the model clone the reference voice and generate speech in that language?
    \end{itemize}
    In previous studies~\cite{zhang2023speak, le2023voicebox}, the language of the reference speaker and the generated language by the model should both exist in great quantity in the MSML dataset. But what if neither of them exist?
    \item  How to realize super-fast speed real-time inference without downgrading the quality, which is crucial for massive commercial production environment.
\end{itemize}

To address the first two problems, OpenVoice is designed to decouple the components in a voice as much as possible. The generation of language, tone color, and other important voice features are made independent of each other, enabling flexible manipulation over individual voice styles and language types. This is achieved without labeling any voice style in the MSML training set. We would like to clarify that \textbf{the zero-shot cross-lingual task in this study is different from that in VALLE-X}~\cite{zhang2023speak}. In VALLE-X, data for all languages need to be included in the MSML training set, and the model cannot generalize to an unseen language outside the MSML training set. By comparison, OpenVoice is designed to generalize to completely unseen languages outside the MSML training set. The third problem is addressed by default, since the decoupled structure reduces requirement on model size and computational complexity. We do not require a large model to learn everything. Also, we avoid auto-regressive or diffusion components to speed up the inference. 

Our internal version of OpenVoice before this public release has been used tens of millions of times by users worldwide between May and October 2023. It powers the instant voice cloning backend of MyShell.ai and has witnessed several hundredfold user growth on this platform. To facilitate the research progress in the field, we explain the technology in great details and make the source code with model weights publicly available.

\begin{figure}
    \centering
    \includegraphics[width=\linewidth]{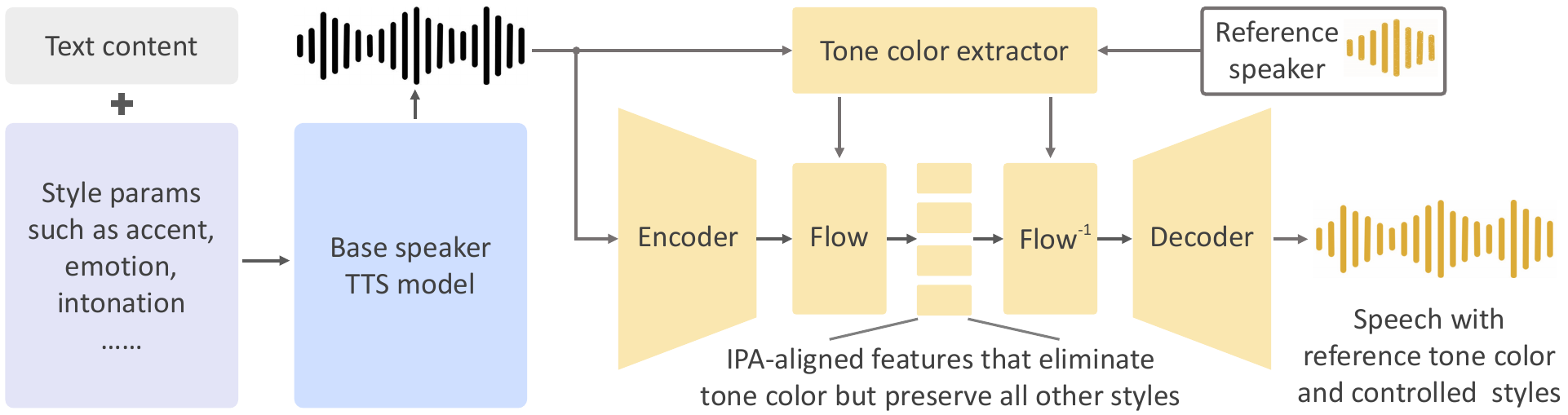}
    \caption{Illustration of the OpenVoice framework. We use a base speaker model to control the styles and languages, and a converter to embody the tone color of the reference speaker into the speech.}
    \label{fig:OpenVoice}
\end{figure}

\section{Approach}
The technical approach is \textbf{simple} to implement but surprisingly effective. We first present the intuition behind OpenVoice, then elaborate on the model structure and training.

\subsection{Intuition}

\textbf{The Hard.} It is obvious that \textbf{simultaneously} cloning the tone color for any speaker, enabling flexible control of all other styles, and adding new language with little effort could be very challenging. It requires a huge amount of combinatorial datasets where the controlled parameters intersect, and pairs of data that only differ in one attribute, and are well-labeled, as well as a relatively large-capacity model to fit the dataset. 

\textbf{The Easy.} We also notice that in regular single-speaker TTS, as long as voice cloning is not required, it is relatively easy to add control over other style parameters and add a new language. For example, recording a single-speaker dataset with 10K short audio samples with labeled emotions and intonation is sufficient to train a single-speaker TTS model that provides control over emotion and intonation. Adding a new language or accent is also straightforward by including another speaker in the dataset.

The intuition behind OpenVoice is to decouple the IVC task into separate subtasks where every subtask is much easier to achieve compared to the coupled task. The cloning of tone color is fully decoupled from the control over all remaining style parameters and languages. \textbf{We propose to use a base speaker TTS model to control the style parameters and languages, and use a tone color converter to embody the reference tone color into the generated voice.}

\subsection{Model Structure}

We illustrate the model structure in Figure.~\ref{fig:OpenVoice}. The two main components of OpenVoice are the base speaker TTS model and the tone color converter. The base speaker TTS model is a single-speaker or multi-speaker model, which allows control over the style parameters (e.g., emotion, accent, rhythm, pauses and intonation), accent and language. The voice generated by this model is passed to the tone color converter, which changes the tone color of the base speaker into that of the reference speaker.

\textbf{Base Speaker TTS Model.} The choice of the base speaker TTS model is very flexible. For example, the VITS~\cite{kim2021conditional} model can be modified to accept style and language embedding in its text encoder and duration predictor. Other choices such as InstructTTS~\cite{yang2023instructtts} can also accept style prompts. It is also possible to use commercially available (and cheap) models such as Microsoft TTS, which accepts speech synthesis markup language (SSML) that specifies the emotion, pauses and articulation. One can even skip the base speaker TTS model, and read the text by themselves in whatever styles and languages they desire. In our OpenVoice implementation, we used the VITS~\cite{kim2021conditional} model by default, but other choices are completely feasible. We denote the outputs of the base model as $\mathbf{X}(L_I, S_I, C_I)$, where the three parameters represent the language, styles and tone color respectively. Similarly, the speech audio from the reference speaker is denoted as $\mathbf{X}(L_O, S_O, C_O)$.

\textbf{Tone Color Converter.} The tone color converter is an encoder-decoder structure with a invertible normalizing flow~\cite{rezende2015variational} in the middle. The encoder is an 1D convolutional neural network that takes the short-time Fourier transformed spectrum of $\mathbf{X}(L_I, S_I, C_I)$ as input. All convolutions are single-strided. The feature maps outputted by the encoder are denoted as $\mathbf{Y}(L_I, S_I, C_I)$. The tone color extractor is a simple 2D convolutional neural network that operates on the mel-spectrogram of the input voice and outputs a single feature vector that encodes the tone color information. We apply it on $\mathbf{X}(L_I, S_I, C_I)$ to obtain vector $\mathbf{v}(C_I)$, then apply it on $\mathbf{X}(L_O, S_O, C_O)$ to obtain vector $\mathbf{v}(C_O)$.

The normalizing flow layers take $\mathbf{Y}(L_I, S_I, C_I)$ and $\mathbf{v}(C_I)$ as input and outputs a feature representation $\mathbf{Z}(L_I, S_I)$ that eliminates the tone color information but preserves all remaining style properties. \textbf{The feature $\mathbf{Z}(L_I, S_I)$ is aligned with International Phonetic Alphabet (IPA)~\cite{international1999handbook} along the time dimension.} Details about how such feature representation is learned will be explained in the next section. Then we apply the normalizing flow layers in the inverse direction, which takes $\mathbf{Z}(L_I, S_I)$ and $\mathbf{v}(C_O)$ as input and outputs $\mathbf{Y}(L_I, S_I, C_O)$. This is a critical step where the tone color $C_O$ from the reference speaker is embodied into the feature maps. Then the $\mathbf{Y}(L_I, S_I, C_O)$ is decoded into raw waveforms $\mathbf{X}(L_I, S_I, C_O)$ by HiFi-Gan~\cite{kong2020hifi} that contains a stack of transposed 1D convolutions. The entire model in our OpenVoice implementation is feed-forward without any auto-regressive component. The tone color converter is conceptually similar to voice conversion~\cite{van2022comparison, polyak2021speech}, but with different emphasis on its functionality, inductive bias on its model structure and training objectives. The flow layers in the tone color converter are structurally similar to the flow-based TTS methods~\cite{kim2021conditional, kim2020glow} but with different functionalities and training objectives.

\hypertarget{choice_of_converter}{\textbf{Alternative Ways and Drawbacks.}}  Although there are alternative ways~\cite{hsu2021hubert, li2023freevc, van2022comparison} to extract $\mathbf{Z}(L_I, S_I)$, we empirically found that the proposed approach achieves the best audio quality. One can use HuBERT~\cite{hsu2021hubert} to extract discrete or continuous acoustic units~\cite{van2022comparison} to eliminate tone color information, but we found that such method also eliminates emotion and accent from the input speech. When the input is an unseen language, this type of method also has issues preserving the natural pronunciation of the phonemes. We also studied another approach~\cite{li2023freevc} that carefully constructs information bottleneck to only preserve speech content, but we observed that this method is unable to completely eliminate the tone color.

\textbf{Remark on Novelty.} OpenVoice does not intend to \textit{invent} the submodules in the model structure. Both the base speaker TTS model and the tone color converter borrow the model structure from existing work~\cite{kim2020glow, kim2021conditional}. The contribution of OpenVoice is the \textbf{decoupled framework that seperates the voice style and language control from the tone color cloning.} This is very simple, but very effective, especially when one wants to control styles, accents or generalize to new languages. If one wanted to have the same control on a coupled framework such as XTTS~\cite{coquixtts}, it could require a tremendous amount of data and computing, and it is relatively hard to fluently speak every language. In OpenVoice, as long as the single-speaker TTS speaks fluently, the cloned voice will be fluent. Decoupling the generation of voice styles and language from the generation of tone color is the core philosophy of OpenVoice. We also provided our \hyperlink{choice_of_converter}{insights} of using flow layers in tone color converter, and the \hyperlink{choice_of_ipa}{importance} of choosing a universal phoneme system in language generalization in our experiment section.

\subsection{Training} \label{sec:training}
In order to train the base speaker TTS model, we collected audio samples from two English speakers (American and British accents), one Chinese speaker and one Japanese speaker. There are 30K sentences in total, and the average sentence length is 7s. The English and Chinese data has emotion classification labels. We modified the VITS~\cite{kim2021conditional} model and input the emotion categorical embedding, language categorical embedding and speaker id into the text encoder, duration predictor and flow layers. The training follows the standard procedure provided by the authors of VITS~\cite{kim2021conditional}. The trained model is able to change the accent and language by switching between different base speakers, and read the input text in different emotions. We also experimented with additional training data and confirmed that rhythm, pauses and intonation can be learned in exactly the same way as emotions.

In order to train the tone color converter, we collected 300K audio samples from 20K individuals. Around 180K samples are English, 60K samples are Chinese and 60K samples are Japanese. This is what we called the MSML dataset. The training objectives of the tone color converter is two-fold. First, we require the encoder-decoder to produce natural sound. During training, we feed the encoder output directly to the decoder, and supervised the generated waveform using the original waveform with mel-spectrogram loss and HiFi-GAN~\cite{kong2020hifi} loss. We will not detail here as it has been well explained by previous literature~\cite{kong2020hifi, kim2021conditional}.

Second, we require flow layers to eliminate as much tone color information as possible from the audio features. During training, for each audio sample, its text is converted to a sequence of \textbf{phonemes in IPA}~\cite{international1999handbook}, and each phoneme is represented by a learnable vector embedding. The sequence of vector embedding is passed to a transformer~\cite{vaswani2017attention} encoder to produce the feature representation of the text content. Denote this feature as $\mathbf{L}\in\mathbb{R}^{c\times l}$, where $c$ is the number of feature channels and $l$ is the number of phonemes in the input text. The audio waveform is processed by the encoder and flow layers to produce the feature representation $\mathbf{Z}\in\mathbb{R}^{c\times t}$, where $t$ is the length of the features along the time dimension. Then we align $\mathbf{L}$ with $\mathbf{Z}$ along the time dimension using dynamic time warping~\cite{senin2008dynamic, muller2007dynamic} (an alternative is monotonic alignment~\cite{kim2020glow, kim2021conditional}) to produce $\mathbf{\bar{L}}\in \mathbb{R}^{c\times t}$, and minimize the KL-divergence between $\mathbf{\bar{L}}$ and $\mathbf{Z}$. Since $\mathbf{\bar{L}}$ does not contain any tone color information, the minimization objective would encourage the flow layers to remove tone color information from their output $\mathbf{Z}$. The flow layers are conditioned on the tone color information from the tone color encoder, which further helps the flow layers to identify what information needs to be eliminated. In addition, we do not provide any style or language information for the flow layers to condition on, which prevents the flow layers to eliminate information other than tone color. Since the flow layers are invertible, conditioning them on a new piece of tone color information and running its inverse process can add the new tone color back to the feature representations, which are then decoded to the raw waveform with the new tone color embodied.

\section{Experiment}

The evaluation of voice cloning is hard to be objective for several reasons. First, different research studies (e.g., \cite{le2023voicebox}, \cite{casanova2022yourtts}) usually have different training and test sets. The numerical comparison could be intrinsically unfair. Even their metrics such as Mean Opinion Score can be evaluated by crowd-sourcing, the diversity and difficulty of the test set would significantly influence the results. For example, if many samples in the test set are neural voices that concentrate on the mean of human voice distributions, then it is relatively easy for most methods to achieve good voice cloning results. Second, different studies usually have different training sets, where the scale and diversity would have considerable influence of the results. Third, different studies can have different focus on their core functionalities. OpenVoice mainly aims at tone color cloning, flexible control over style parameters, and making cross-lingual voice clone easy even without massive-speaker data for a new language. These are different from the objectives of previous work on voice cloning or zero-shot TTS. Therefore, instead of comparing numerical scores with existing methods, we mainly focus on analyzing the qualitative performance of OpenVoice itself, and make the audio samples publicly available for relevant researchers to freely evaluate.

\textbf{Accurate Tone Color Cloning.} We build a test set of reference speakers selected from celebrities, game characters and anonymous individuals. The test set covers a wide voice distributions including both expressive unique voices and neutral samples in human voice distribution. With any of the 4 base speakers and any of the reference speaker, OpenVoice is able to accurately clone the reference tone color and generate speech in multiple languages and accents. We invite the readers to this website\footnote{https://research.myshell.ai/open-voice/accurate-tone-color-cloning} for qualitative results.

\textbf{Flexible Control on Voice Styles.} A premise for the proposed framework to flexibly control the speech styles is that the tone color converter is able to only modify the tone color and preserves all other styles and voice properties. In order to confirm this, we use both our base speaker model and the Microsoft TTS with SSML to generate a speech corpus of 1K samples with diverse styles (emotion, accent, rhythm, pauses and intonation) as the base voices. After converting to the reference tone color, we observed that all styles are well-preserved. In rare cases, the emotion will be slightly neutralized, and one way that we found to solve this problem is to replace the tone color embedding vector of this particular sentence with the average vector of multiple sentences with different emotions from the same base speaker. This gives less emotion information to the flow layers so that they do not eliminate the emotion. Since the tone color converter is able to preserve all the styles from the base voice, controlling the voice styles becomes very straightforward by simply manipulating the base speaker TTS model. The qualitative results are publicly available on this website\footnote{https://research.myshell.ai/open-voice/flexible-voice-style-control}.

\textbf{Cross-Lingual Voice Clone with Ease.} OpenVoice achieves near zero-shot cross-lingual voice cloning without using any massive-speaker data for an unseen language. It does require a base speaker of the language, which can be achieved with minimum difficulty with the off-the-shelf models and datasets. On our website\footnote{https://research.myshell.ai/open-voice/zero-shot-cross-lingual-voice-cloning}, we provide an abundant of samples that demonstrates the cross-lingual voice clone capabilities of the proposed approach. The cross-lingual capabilities are two-fold:
\begin{itemize}
    \item When the language of the reference speaker is unseen in the MSML dataset, the model is able to accurately clone the tone color of the reference speaker.
    \item When the language of the generated speech is unseen in the MSML dataset, the model is able to clone the reference voice and speak in that language, as long as the base speaker TTS supports that language.
\end{itemize}

\textbf{Fast Inference with Low Cost.} Since OpenVoice is a feed-forward structure without any auto-regressive component, it achieves very high inference speed. Our experiment shows that a slightly optimized version of OpenVoice (including the base speaker model and the tone converter) is able achieve $12\times$ real-time performance on a single A10G GPU, which means it only takes 85ms to generate a one second speech. Through detailed GPU usage analysis, we estimate that the upper bound is around $40\times$ real-time, but we will leave this improvement as future work.

\hypertarget{choice_of_ipa}{\textbf{Importance of IPA.}} We found that using IPA as the phoneme dictionary is crucial for the tone color converter to perform cross-lingual voice cloning. As we detailed in Section~\ref{sec:training}, in training the tone color converter, the text is first converted into a sequence of phonemes in IPA, then each phoneme is represented by a learnable vector embedding. The sequence of embedding is encoded with transformer layers and compute loss against the output of the flow layers, aiming to eliminate the tone color information. IPA itself is a cross-lingual unified phoneme dictionary, which enables the flow layers to produce a language-neutral representation. Even if we input a speech audio with unseen language to the tone color converter, it is still able to smoothly process the audio. We also experimented with other types of phoneme dictionaries but the resulting tone color converter tend to mispronounce some phonemes in unseen languages. Although the input audio is correct, there is a high likelihood that the output audio is problematic and sounds non-native.

\section{Discussion}
OpenVoice demonstrates remarkable instance voice cloning capabilities and is more flexible than previous approaches in terms of voice styles and languages. The intuition behind the approach is that it is relatively easy to train a base speaker TTS model to control the voice styles and languages, as long as we do not require the model to have the ability to clone the tone color of the reference speaker. Therefore, we proposed to decouple the tone color cloning from the remaining voice styles and the language, which we believe is the foundational design principle of OpenVoice. In order to facilitate future research, we make the source code and model weights publicly available.

\bibliographystyle{ieee}
\bibliography{reference}

\end{document}